\documentclass{ws-ijmpd}
\usepackage{graphicx}
\begin{document}

\markboth{Parthapratim Pradhan}{(Horizon Straddling ISCOs in Spherically Symmetric String Black Holes)}

\title{Horizon Straddling ISCOs in Spherically Symmetric String Black Holes }
\author{Parthapratim Pradhan\footnote{pppradhan77@gmail.com.}}

\address{Department of Physics, Vivekananda Satavarshiki Mahavidyalaya(Affiliated to Vidyasagar University),
Manikpara, Jhargram, West Midnapur, West Bengal~721513, India}

\maketitle

\begin{history}
\received{Day Month Year}
\revised{Day Month Year}
\comby{Managing Editor}
\end{history}

\begin{abstract}
The causal geodesics in the equatorial plane of a static  extremal charged black holes in heterotic string theory 
are examined with regard to their geodesic stability, and compared with similar geodesics in the non-extremal situation. 
Extremization of the effective potential for time-like  and null circular geodesics implies that  in the extremal 
limit, the radius of ISCO(Inner-most Stable Circular Orbit) $(r_{ISCO})$, circular photon orbit (CPO) $(r_{ph})$ 
and marginally bound circular orbit (MBCO) $(r_{mb})$ \emph{are coincident with the event horizon $(r_{hor})$ i.e. 
$r_{ISCO}=r_{ph}=r_{mb}=r_{hor}=2M$}. Since the proper radial distance on a constant time slice both in 
Schwarzschild and Painlev\'{e}-Gullstrand coordinates become zero, thus these three orbits indeed 
coincide with the \emph{null geodesic generators of the event horizon}. This strange behavior is quite 
different from the static, spherically symmetric extremal Reissner Nordstr{\o}m  black hole. 
\end{abstract}


\section{Introduction}
The  motion of geodesics determine important features of the black hole(BH) spacetime\cite{sch}. Among the 
different kinds of geodesics, circular geodesics specially, ISCOs are more interesting. Studies of neutral 
test particles, both time-like and null, is one way to understand the gravitational field around a BH 
spacetime. Different theoretical as well as observational predictions i.e. the gravitational red-shift, 
gravitational bending of light, Shapiro time-delay, Lense-Thirring precession, gravitational lensing 
and perihelion shift etc. all are physical phenomenon which might be related to the geodesic structure 
of the BH\cite{hart}. Thus it is important to study the proper geodesic structure of the BH spacetime.

It is well known that ISCO [also called marginally stable circular orbit(MSCO)] or LSCO(Last stable 
circular orbit) plays a crucial role in accretion disk thory. It also useful to estimate the 
temperature of the accretion disk(ad) via the Eddington luminosity \cite{hart} as 
\begin{eqnarray}
 T_{ad} \sim 1 \times 10^{8} \left(\frac{GM}{c^{2}R}\right)^{1/2}\left(\epsilon \frac{M_{\odot}}{M}\right)^{1/4} K \nonumber \\
                    \sim 9 \left(\frac{GM}{c^{2}R}\right)^{1/2}\left(\epsilon \frac{M_{\odot}}{M}\right)^{1/4} KeV.
\end{eqnarray}
It has been observed that for neutron star's surface $\frac{GM}{c^{2}R} \sim 0.1$, for Schwarzschild BH 
$\frac{GM}{c^{2}R} \sim \frac 1 6$ and we find for extremal string BH  $\frac{GM}{c^{2}R} \sim \frac{1}{2}$. 
Here $R$ is the radius of ISCO. In any cases, for $M \sim M_{\odot}$, $\epsilon \sim 0.5$, we find the 
characteristic temperature of the X-ray sources around $T_{ad} \sim $ few KeV. To compute the binding 
energy, ISCO also plays an important role \cite{st}. 

In Einstein's general relativity, circular geodesics of arbitrary radii are not possible, there exists a minimum
radii below which no circular orbits are possible. The conditions for the existence of ISCO, MBCO  and CPO have 
been considered for the Schwarzschild black hole \cite{sch}, Reissner Nordstr{\o}m(RN) black-hole \cite{sch,pp1,pqr}, 
spherically symmetric string BH \cite{bb,fer} and charged dilation BH\cite{ms}. In \cite{fer}, the author completely 
studied the null geodesics of charged BHs in string theory. Gravitational lensing effect of the charged string BHs 
examined in \cite{ab}.

But the study of causal geodesics of charged black holes in string theory, both \emph{extreme} and non-extremal 
situations have not been considered in the extant literature. The fact that string theory is a promising 
candidate for a consistent quantum theory of gravity. Hence, the studies of BH solution of the low energy 
string theory has a great significance from this perspective. Additionally, the classical equation of motion 
for string theory has the form of Einstein's equation plus the Planck scale corrected terms. 

Thus in the present work, 
we wish to investigate in detail the equatorial time-like circular geodesics and null circular
geodesics, both extremal and non-extremal cases of four dimensional static, spherically symmetric  string BH\cite{gm,ghs}, 
which is also called Gibbons-Maeda-Garfinkle-Horowitz-Strominger (GMGHS) BH and we show that in the \emph{extremal limit}, 
the radius of ISCO, CPO and MBCO coalesces to the same point i.e.
\begin{eqnarray}
r_{ISCO}=r_{ph}=r_{mb}=r_{hor}=2M.~\label{rel}
\end{eqnarray}
And we also compute that the proper radial distances on a constant time slice, both in Schwarzschild and
Painlev\'{e}-Gullstrand\cite{pain} coordinates, from the horizon to the ISCO becomes zero, therefore
these three orbits indeed coincide \emph{with the principal null geodesics generators} of the event 
horizon. From the best of my knowledge this result is not reported previously in the literature.

The paper is organized as follows. In section 2, we give the basics of GMGHS space-time. In section 3, we
shall analyze in  detail the equatorial circular geodesics, both particle orbits and photon orbits for non
extreme GMGHS space-time and also compute the ISCO for non-extreme GMGHS BH.
In section 4, we present the  particle orbits and photon orbits for extreme GMGHS space-time and compute
the ISCO for extreme GMGHS BH. Discussion of the proper radial distance is presented in section 5. Implications 
of different useful coordinates for GMGHS space-time are described shortly in section 6.  Finally, the key 
conclusions are given in section 7.

\section{Preliminaries of the GMGHS Space-time:}

The effective action in heterotic string theory\cite{as,horo} in the low energy limit is represented by
\begin{eqnarray}
S &=& \frac{1}{16\pi} \int d^{4}x\sqrt{-g}[R-\frac{1}{12}e^{-4\Phi}H_{abc}H^{abc}-2(\nabla\Phi)^{2}-e^{-2\Phi}
F_{ab}F^{ab}]
\end{eqnarray}
where $g_{ab}$ is the metric, $\Phi$ is the dilation field, $R$ is the scalar curvature
and $F_{ab}=\partial_{a}A_{b} -\partial_{b}A_{a}$ is the field strength corresponds to
the Maxwell field $A_{a}$, and $F_{ab}$ is the Maxwell field associated with a $U(1)$
subgroup of $E_{8}\times E_{8}$ or $\frac{spin(32)}{Z_{2}}$, and
\begin{eqnarray}
H_{abc} &=& \partial_{a}B_{bc}+ \partial_{b}B_{ca}+\partial_{c}B_{ab}-(\Omega_{3}(A))_{abc}
\end{eqnarray}
where $B_{ab}$ is the antisymmetric tensor gauge field, and
\begin{eqnarray}
(\Omega_{3}(A))_{abc} &=& \frac{1}{4}\left(A_{a}F_{bc}+A_{b}F_{ca}+A_{c}F_{ab} \right)
\end{eqnarray}
is the gauge Chern-Simons term. We are interested to explore in this work to the situation when the
fields $H_{abc}$ and $B_{ab}$ are corresponds to the zero value. Therefore the above action reduces to
\begin{eqnarray}
S &=& \frac{1}{16\pi} \int d^{4}x\sqrt{-g}[R-2(\nabla\Phi)^{2}-e^{-2\Phi}F_{ab}F^{ab}]
\end{eqnarray}
and the corresponding field equations are
\begin{eqnarray}
\nabla_{a}(e^{-2\Phi}F^{ab})   &=& 0 \\
\nabla^{2}\Phi+\frac{1}{2}e^{-2\Phi}F^{2} &=& 0
\end{eqnarray}
\begin{eqnarray}
R_{ab}&=& -2\nabla_{a}\Phi\nabla_{b}\Phi-2e^{-2\Phi}F_{ac}F^{c}_{b}+\frac{1}{2}g_{ab}e^{-2\Phi}F^{2}
\end{eqnarray}

The static charged BH solutions of the above action was found by Gibbons and Maeda \cite{gm} in 1988,
and independently by Garfinkle, Horowitz and Strominger \cite{ghs} in 1991. 

The line element of this charged BH is given by
\begin{eqnarray}
ds^2 = -\left(1-\frac{2M}{r}\right)dt^{2}+\left(1-\frac{2M}{r}\right)^{-1}dr^{2}+r\left(r-b\right)
\left(d\theta^{2}+\sin^{2}\theta d\phi^{2}\right) ~.\label{gh}
\end{eqnarray}
where, $b=\frac{Q^{2}}{M}e^{-2\phi_{0}}$
and
\begin{eqnarray}
e^{-2\phi} &=& e^{-2\phi_{0}}\left(1-\frac{b}{r}\right),\,\,
\mbox{and}\,\, F = Q\sin\theta d\theta\wedge d\phi
\end{eqnarray}
where $\phi_{0}$ is the asymptotic value of the dilation field, $M$ represents the mass of the BH, $Q$ denote
its electric charge and $\phi$ is the scalar field. We have considered throughout the work when the strength
of the dilation field is precisely zero i.e. $\phi_{0}=0$. The BH has a regular event horizon at $r_{hor}=2M$,
which is  identical to the Schwarzschild BH. There are some important differences between RN BH and
GMGHS BH which are tabulated  as below:

\begin{center}
\begin{tabular}{|c|c|c|}
    \hline
    Properties & RN BH & GMGHS BH \\
    \hline
    Horizon: & $r_{\pm}=M\pm\sqrt{M^{2}-Q^{2}}$ & $r_{hor}=M$ \\
    Extremal limit: & $ M^{2}=Q^{2}$ & $Q^{2}=2M^{2}e^{2\phi_{0}} $  \\
    Area: & $A_{\pm}=4\pi r_{\pm}^{2}$  & ${\cal A}=4\pi r_{hor}(r_{hor}-b)$ \\
    Hawking Temp.: & $T_{H}=\frac{\sqrt{M^{2}-Q^{2}}}{2\pi(M+\sqrt{M^{2}-Q^{2}})}$ & $T_{H}=\frac{e^{-\phi_{0}}}{8\pi M}$ \\
    Area at extremality:&  $A_{ex}=4\pi M^{2}$ & ${\cal A}_{ex}=0$ \\
        \hline
\end{tabular}
\end{center}

\section{Equatorial Circular Geodesics in GMGHS BH:}

Since the space-time has a time-like isometry generated by the time-like Killing vector $\xi \equiv \partial_t$ whose projection
along the 4-velocity ${\bf u}$  of geodesics: $\xi \cdot {\bf u} = -E$, is conserved along such geodesics.
There is also the `angular momentum' $L \equiv \zeta \cdot {\bf u}$  (where $\zeta \equiv \partial_{\phi}$)
which is similarly conserved. Using these properties together with the normalization of the four velocity, one
obtains the radial equation for charged BHs in string theory on the $\theta=\frac{\pi}{2}$ plane:
\begin{eqnarray}
(u^{r})^{2}=\dot{r}^{2}=E^{2}-{\cal V}_{eff}=E^{2}-\left(\frac{L^{2}}{r(r-b)}-\epsilon \right)\left(1-\frac{2M}{r}\right)~.\label{radial}
\end{eqnarray}
where the standard effective potential for GMGHS space-time is
\begin{eqnarray}
{\cal V}_{eff}=\left(\frac{L^{2}}{r(r-b)}-\epsilon \right)\left(1-\frac{2M}{r}\right) ~.\label{vrn}
\end{eqnarray}
Here, $\epsilon=-1$ for time-like geodesics, $\epsilon=0$ for light-like geodesics and $\epsilon=+1$ for space-like geodesics.

\subsection{Particle Orbits:}
\begin{figure}[t]
\begin{center}
{\includegraphics[width=0.45\textwidth]{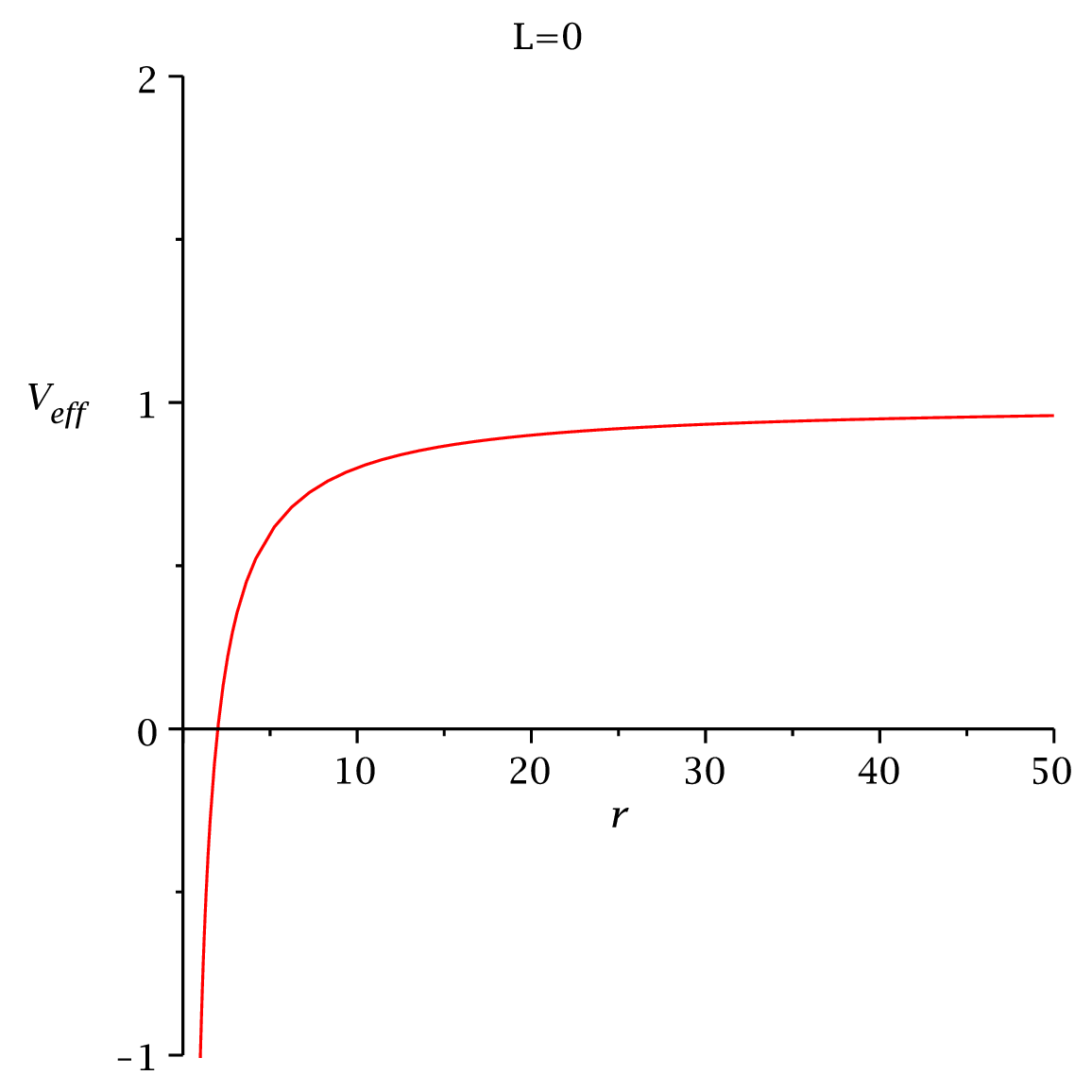}}
\end{center}
\caption{The figure shows the variation  of ${\cal V}_{eff}$  with $r$.  Here, $M =1$. \label{rvef}}
\end{figure}

The effective potential for massive particles could be obtained from the above equation by
substituting $\epsilon=-1$ :

\begin{eqnarray}
{\cal V}_{eff}=\left(1+\frac{L^{2}}{r(r-b)}\right)\left(1-\frac{2M}{r}\right) ~.\label{vr}
\end{eqnarray}

a) When $L=0$, we get the radial geodesics and correspondingly the effective potential becomes
\begin{eqnarray}
{\cal V}_{eff}=1-\frac{2M}{r} ~.\label{trp}
\end{eqnarray}
If we observe the effective potential for radial time-like geodesics graphically it looks like
as in Fig. \ref{rvef}.
%\newpage
In the limit $b=0$ or $Q=0$, we obtain the effective potential for well known Schwarzschild black hole. 
\begin{figure}[t]
\begin{center}
{\includegraphics[width=0.45\textwidth]{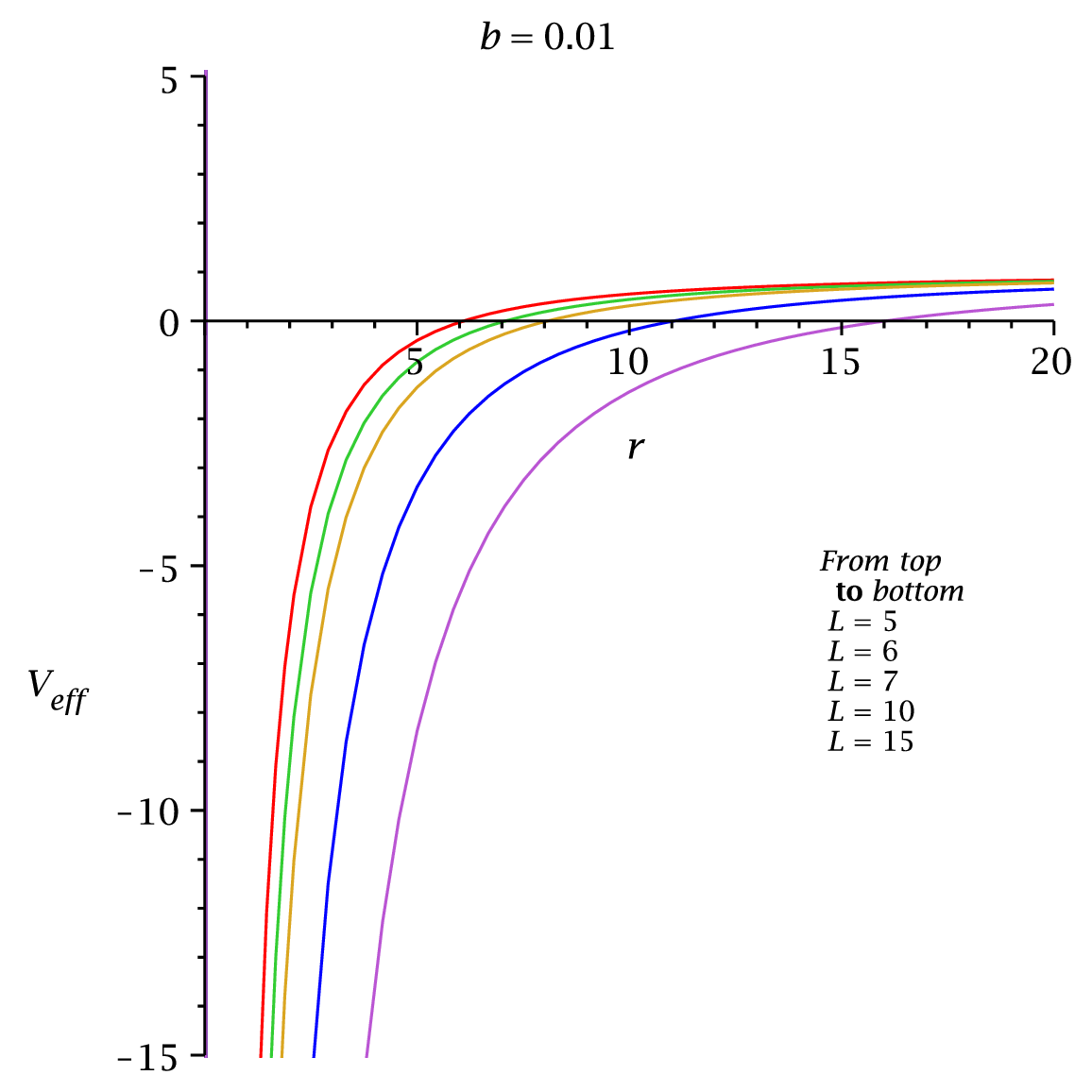}}
{\includegraphics[width=0.45\textwidth]{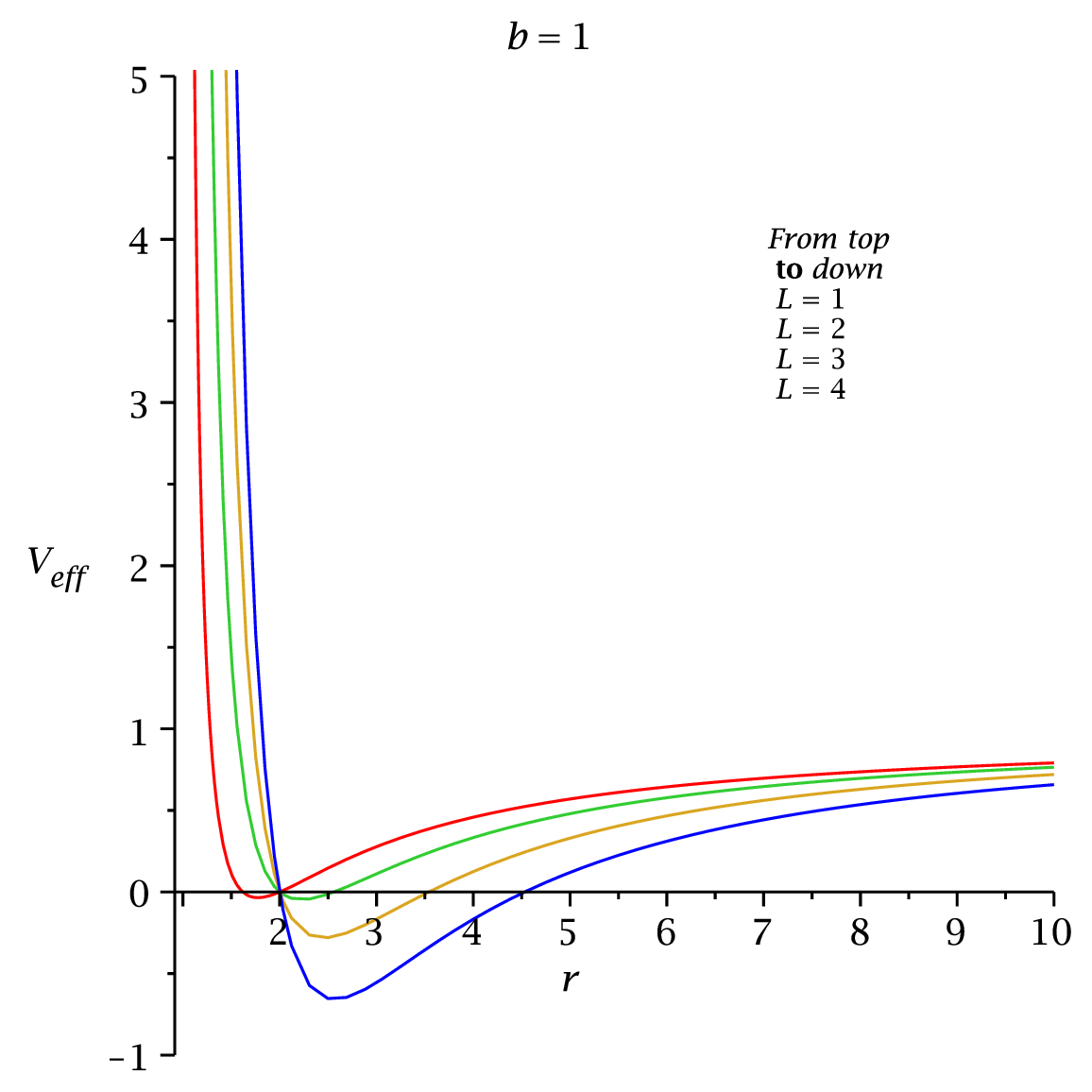}}
\end{center}
\caption{The picture shows the variation  of ${\cal V}_{eff}$  with $r$.  Here, $M =1$. \label{gmcase}}
\end{figure}
\begin{figure}[t]
\begin{center}
{\includegraphics[width=0.45\textwidth]{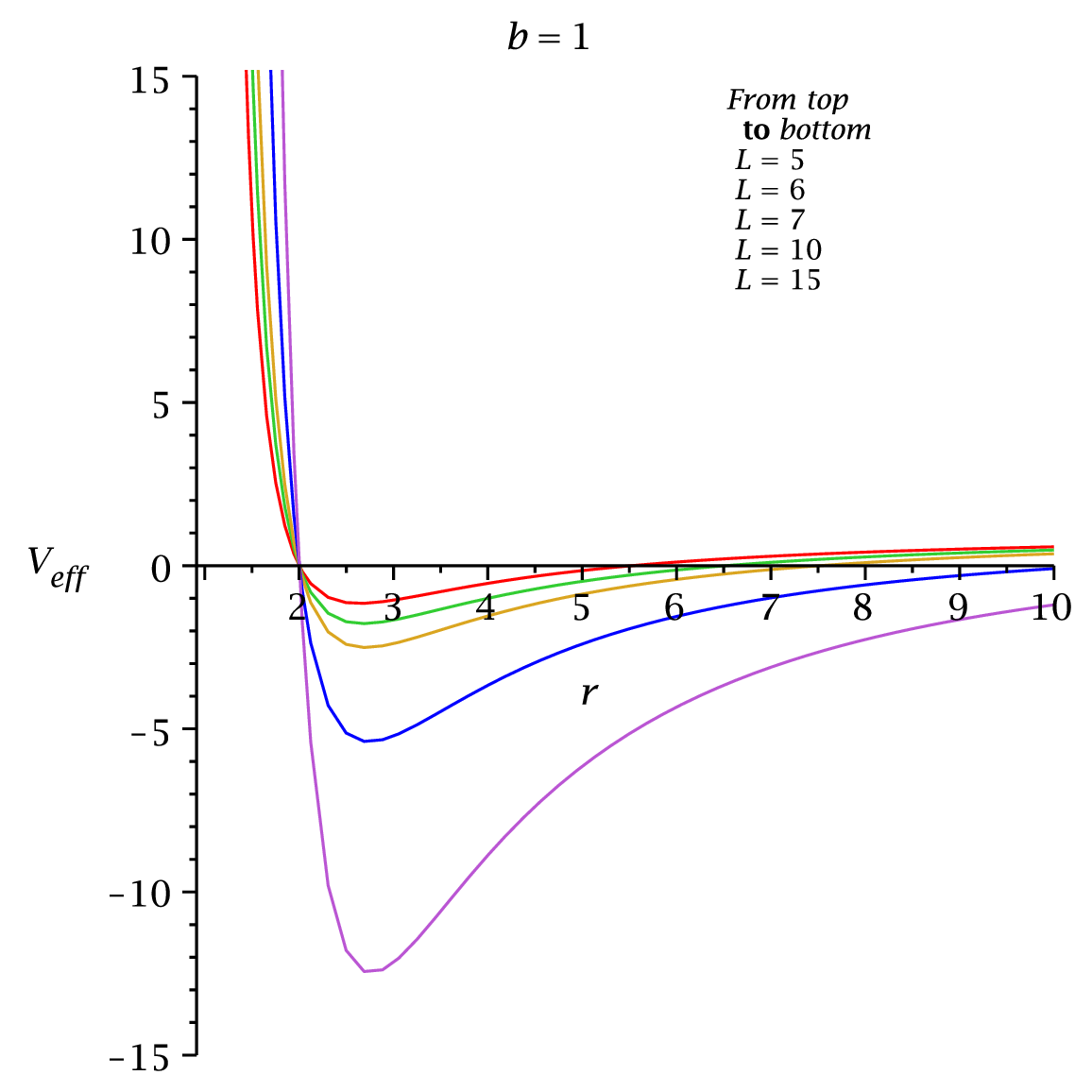}}
\end{center}
\caption{The picture shows the variation  of ${\cal V}_{eff}$  with $r$.  Here, $M =1$.
 \label{gmcase1}}
\end{figure}

In the diagrams ( Fig. \ref{gmcase}, Fig. \ref{gmcase1}), we have described the properties of effective potential
as derived in the Eq. (\ref{vr}) for different values of $b$.

\begin{figure}[t]
\begin{center}
{\includegraphics[width=0.45\textwidth]{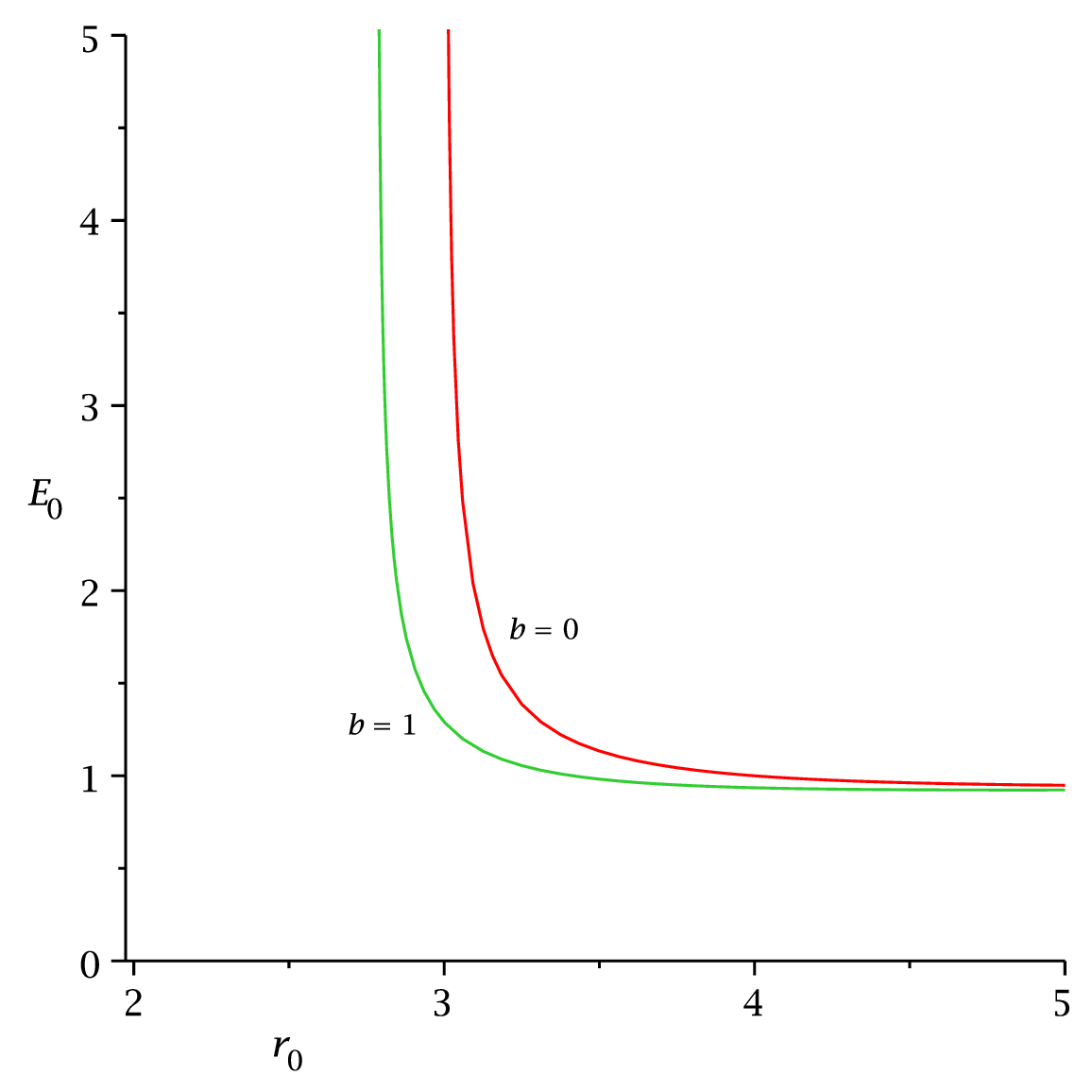}}
{\includegraphics[width=0.45\textwidth]{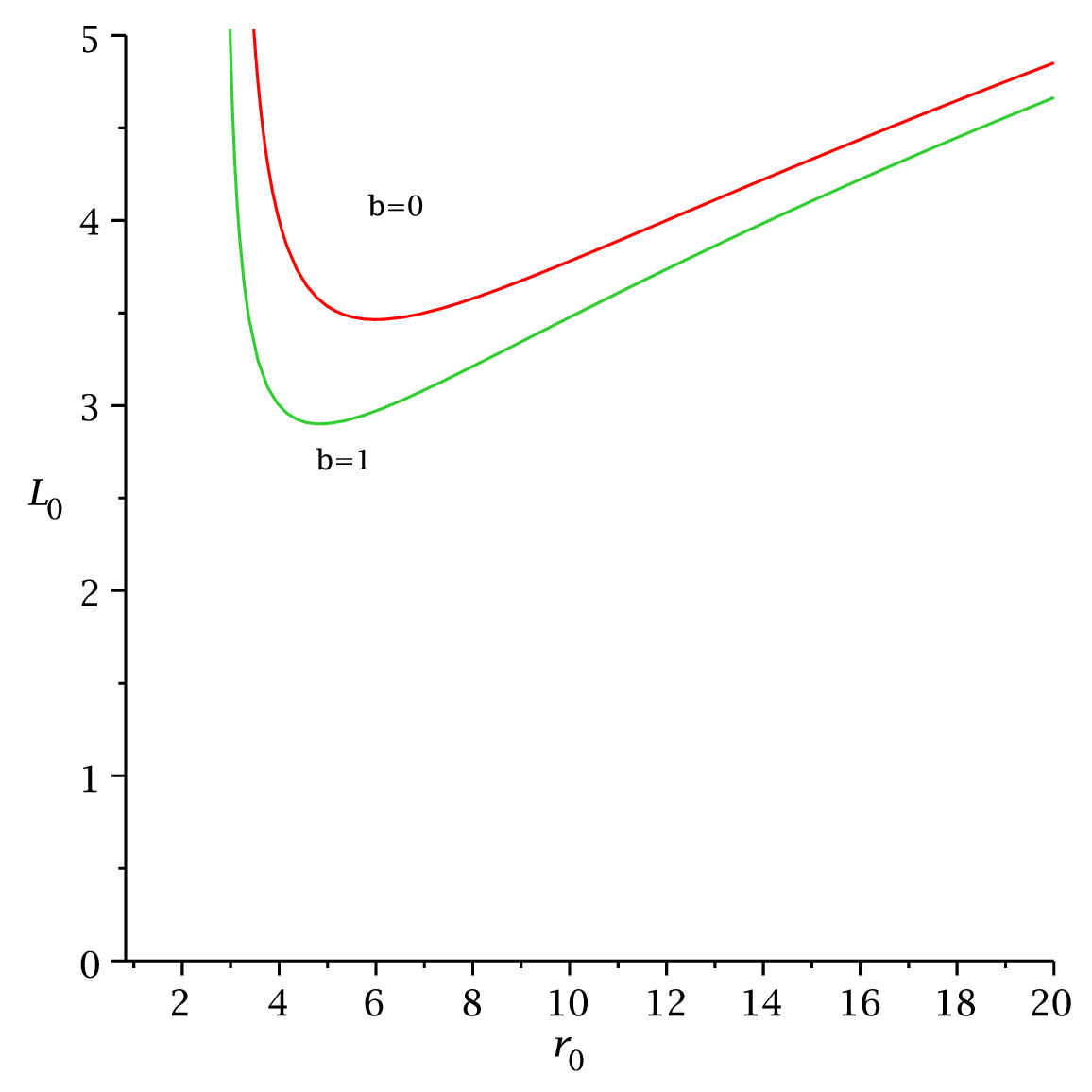}}
\end{center}
\caption{The figures show the variation  of ${E}_{0}$ and ${L}_{0}$  with $r_{0}$.  Here, $M =1$.
\label{el}}
\end{figure}

To compute the circular geodesic motion of the test particle in the Einstein -Maxwell gravitational field , we
must have $r=r_{0}=constant$  and from the equation (\ref{radial}), we get
\begin{eqnarray}
{\cal V}_{eff} &=& E^{2} ~.\label{v}
\end{eqnarray}
and
\begin{eqnarray}
\frac{d{\cal V}_{eff}}{dr} &=& 0 ~.\label{dvdr}
\end{eqnarray}
Therefore we obtain the energy and angular momentum per unit mass of the test particle along the circular orbits are given by

\begin{eqnarray}
E_{0} &=& \sqrt{\frac{(2r_{0}-b)(r_{0}-2M)^{2}}{r_{0}[2r_{0}^{2}-(b+6M)r_{0}+4Mb]}} ~.\label{engg}
\end{eqnarray}

and
\begin{eqnarray}
L_{0} &=& \sqrt{\frac{2Mr_{0}(r_{0}-b)^2}{2r_{0}^{2}-(b+6M)r_{0}+4Mb}} ~ .\label{angg}
\end{eqnarray}
%\newpage
We have plotted the energy and angular momentum of the test particle in  Fig. \ref{el}.

Circular motion of the test particle to be exists when both energy and angular momentum are real and
finite, therefore we must have
$2r_{0}^{2}-(b+6M)r_{0}+4Mb>0$ and $r_{0}>b$.
Again the angular frequency measured by an asymptotic observers for timelike circular geodesics at
$r=r_{0}$  is given by
\begin{equation}
\Omega_{0}=\frac{u^\phi}{u^t}=\sqrt{\frac{2M}{r_{0}^2(2r_{0}-b)}}
\end{equation}

\begin{figure}[t]
\begin{center}
{\includegraphics[width=0.45\textwidth]{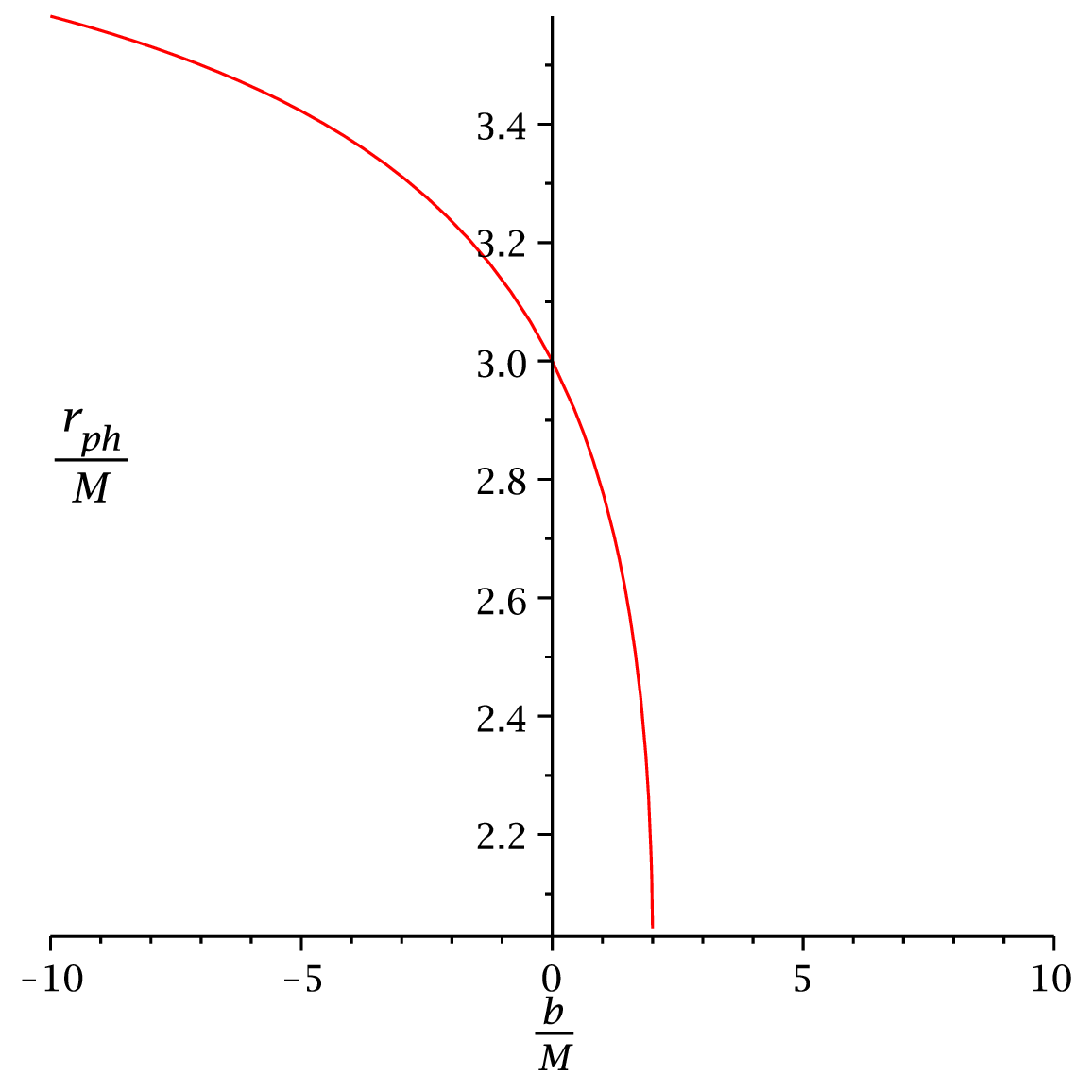}}
\end{center}
\caption{The figure displays the variation  of $\frac{r_{ph}}{M}$
with $\frac{b}{M}$.  Here, $Q=1$. \label{phcase}}
\end{figure}

In the limit $b\rightarrow 0$, we obtain the angular frequency for Schwarzschild BH which is $\Omega_{0}=\sqrt{\frac{M}{r_{0}^3}}$.
In general relativity, circular orbits do not exist for all values of $r$, so the denominator of equations (\ref{engg},\ref{angg})
real only if $2r_{0}^{2}-(b+6M)r_{0}+4Mb\geq 0$. The limiting case of equality gives a circular orbit with indefinite energy per
unit mass, i.e a photon orbit. This photon orbit is the innermost boundary of the circular orbits for massive particles. Comparing
the above equation of particle orbits with (\ref{phr1}) when $r_{0}=r_{c}$, we can see that photon orbits are the limiting case
of time-like circular orbit.  It occurs at the radius
\begin{equation}
 r_{ph} = \frac{1}{4}(b+6M+\sqrt{b^2-20Mb+36M^2})
\end{equation}
It is described in the Fig. \ref{phcase}.

\begin{figure}[t]
\begin{center}
{\includegraphics[width=0.45\textwidth]{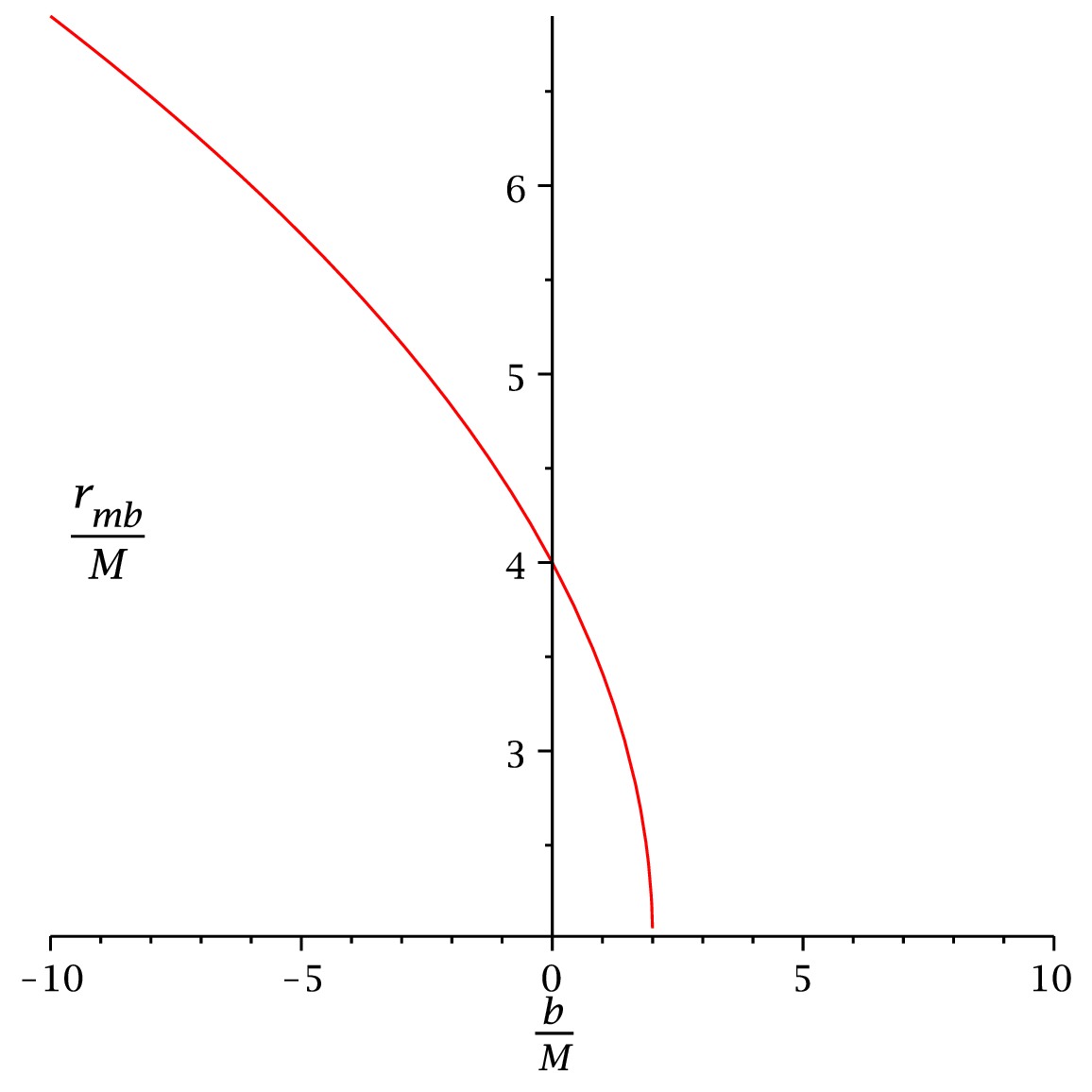}}
\end{center}
\caption{The picture depicts the variation  of $\frac{r_{mb}}{M}$  with $\frac{b}{M}$.  Here, $b=1$. \label{mbcase}}
\end{figure}

The MBCO \cite{sch} can be obtained by setting
$E^{2}=1$, then the radius of MBCO is located at
\begin{eqnarray}
r_{mb} &=& 2M\pm\sqrt{2M(2M-b)}
\end{eqnarray}
It has been found in the Fig. \ref{mbcase}.

In the limit $b\rightarrow 0$, we get $r_{mb}=4M$, which is the radius of
MBCO for Schwarzschild BH. At the extreme limit $b=2M$, it occurs at the radius $r_{mb}=2M$.

From the astrophysical point of view the most important class of orbit is innermost stable circular orbit(ISCO),
which occurs at the point of inflection of the effective potential ${\cal V}_{eff}$. Thus at the point of
inflection
\begin{eqnarray}
\frac{d^2{\cal V}_{eff}}{dr^2} &=& 0 \label{pi}
\end{eqnarray}
with the auxiliary equation $\frac{d{\cal V}_{eff}}{dr}=0$. Then the ISCO equation for GMGHS BH is given by
\begin{eqnarray}
 r_{0}^3-6Mr_{0}^2+6Mbr_{0}-2Mb^2 &=& 0
\end{eqnarray}

\begin{figure}[t]
\begin{center}
{\includegraphics[width=0.45\textwidth]{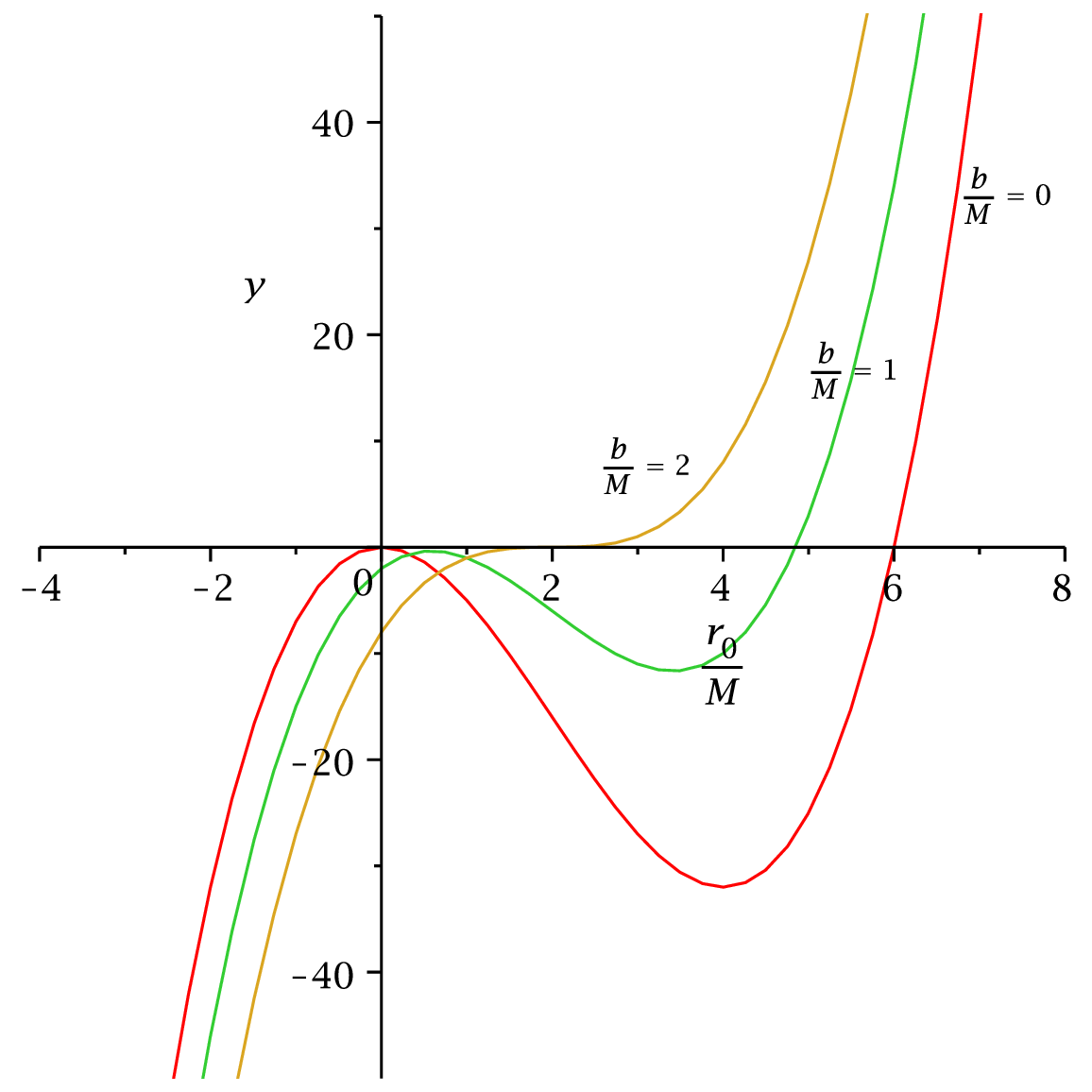}}
\end{center}
\caption{The stability threshold is defined by the largest real root of the cubic
$y=r_{0}^3-6Mr_{0}^2+6Mbr_{0}-2Mb^2=0$. For $b=0$ this root has the value $r_{0}=6M$. For
$b=2M$ it has the value $r_{0}=2M$ .The figure shows the variation
of $y$  with $\frac{r_{0}}{M}$.  \label{y11}}
\end{figure}

\begin{figure}[t]
\begin{center}
{\includegraphics[width=0.45\textwidth]{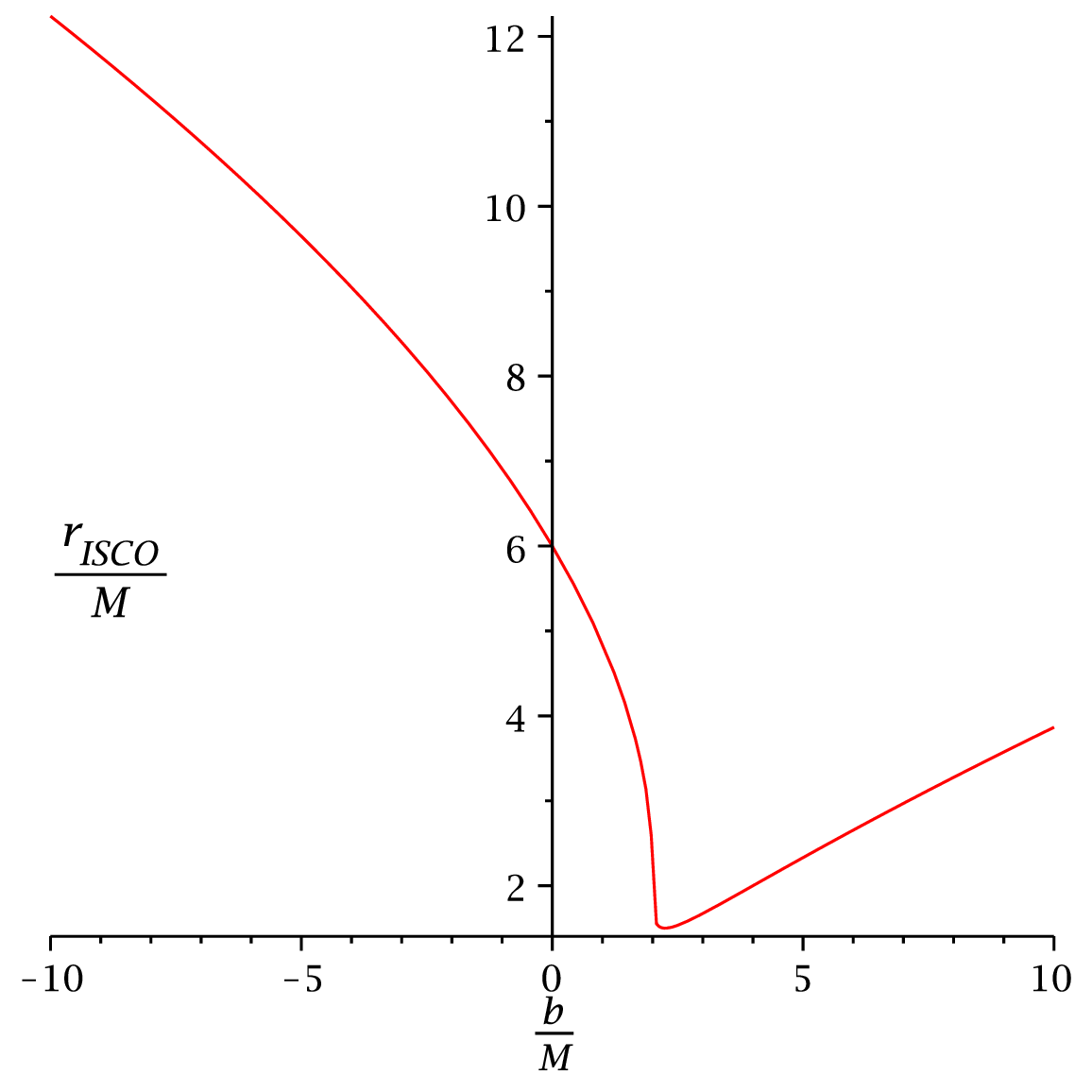}}
\end{center}
\caption{The figure shows the variation  of $\frac{r_{ISCO}}{M}$  with $\frac{b}{M}$.  Here, $b=1$. \label{isccase}}
\end{figure}

The real root of the equation gives the radius of
ISCO at $r_{0}=r_{ISCO}$ which is given by
\begin{eqnarray}
 \frac{r_{ISCO}}{M} &=& 2+Z+\frac{2(2-\frac{b}{M})}{Z}\\
 Z &=& \left[8-6(\frac{b}{M})+(\frac{b}{M})^{2}+
 \sqrt{(\frac{b}{M})^{4}-4(\frac{b}{M})^{3}+4(\frac{b}{M})^{2}}\right]^{\frac{1}{3}}
\end{eqnarray}
It could be found in the Fig. \ref{isccase}.

In the limit $b\rightarrow 0$, we obtain $r_{ISCO}=6M$, which is the radius
of ISCO for Schwarzschild BH.

\subsection{Photon Orbits:}
The radial potential that governs the null geodesics can be expressed as
\begin{eqnarray}
 {\cal U}_{eff}&=& \frac{L^2}{r(r-b)}\left(1-\frac{2M}{r}\right)
\end{eqnarray}
In the limit $b=0$, we get the effective potential for Schwarzschild case.
\newpage
In the following Fig. \ref{gmpcase}, we have drawn the effective potential
for photon orbits of GMGHS BH and for various values of $b$.
\begin{figure}[t]
\begin{center}
{\includegraphics[width=0.45\textwidth]{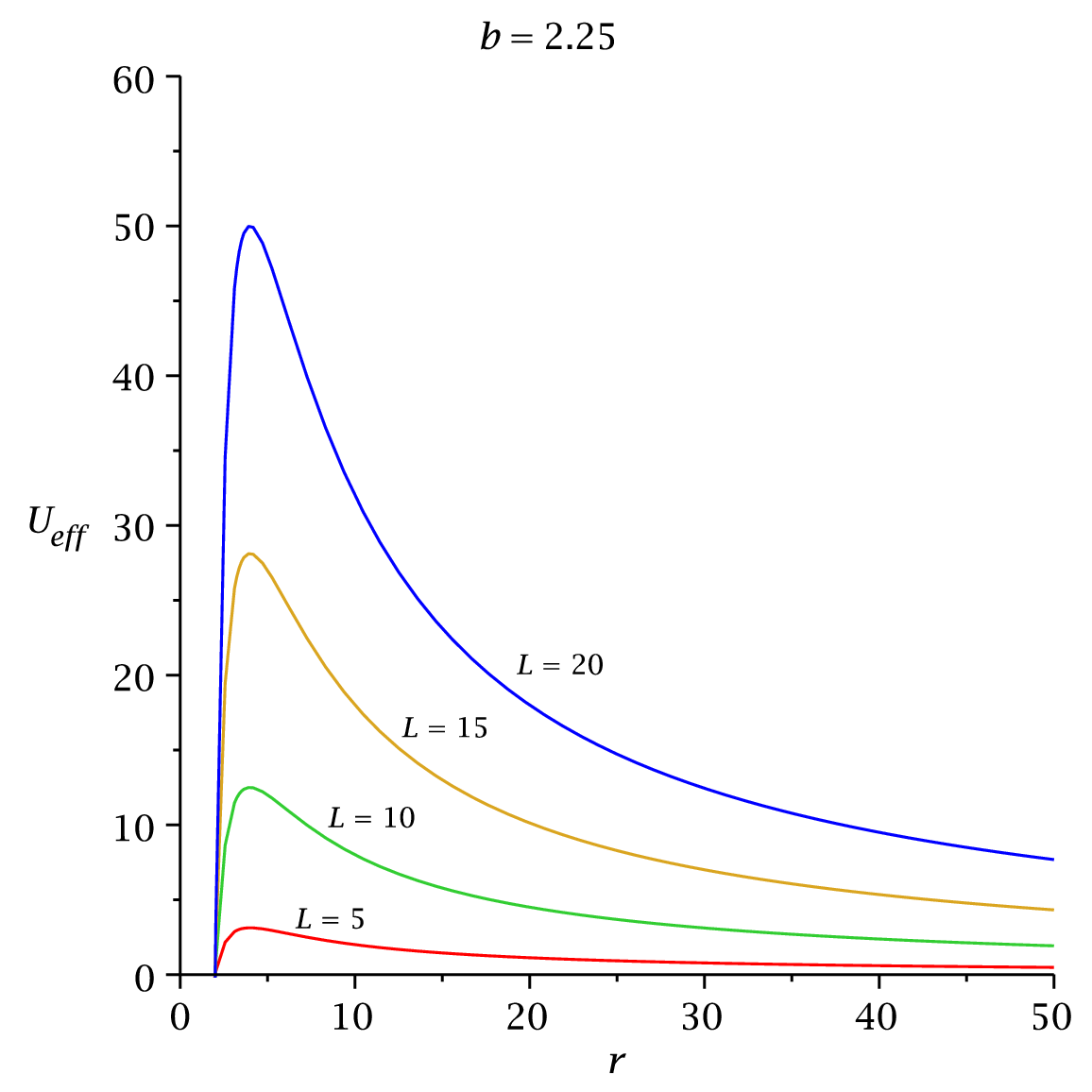}}
{\includegraphics[width=0.45\textwidth]{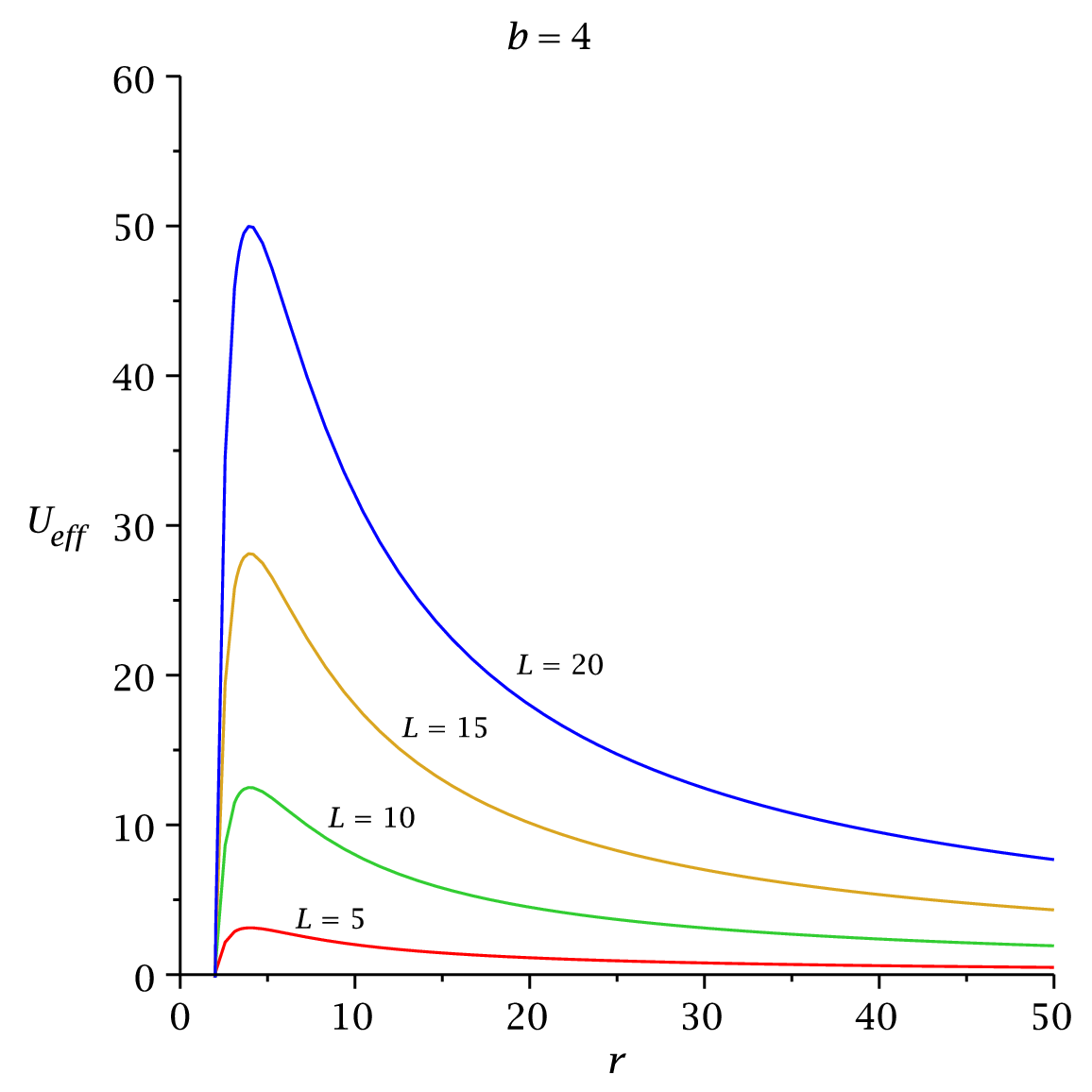}}
\end{center}
\caption{The picture shows the variation  of ${\cal U}_{eff}$  with $r$.  Here, $M =1$.
 \label{gmpcase}}
\end{figure}

For circular null geodesics at $r=r_{c}$:
\begin{eqnarray}
 {\cal U}_{eff} &=& E^2
\end{eqnarray}
and
\begin{eqnarray}
 \frac{d{\cal U}_{eff}}{dr} &=& 0
\end{eqnarray}
Thus we obtain the ratio of energy and angular momentum of the test particle evaluated at $r=r_{c}$ for CPO:
\begin{eqnarray}
 \frac{E_{c}}{L_{c}} &=& \pm \sqrt{\frac{r_{c}-2M}{r_{c}^2(r_{c}-b)}}
\end{eqnarray}
and
\begin{eqnarray}
 2r_{c}^2-(b+6M)r_{c}+4bM &=& 0.~\label{phr1}
\end{eqnarray}
After introducing the impact parameter $D_{c}=\frac{L_{c}}{E_{c}}$, the above equation reduces to
\begin{eqnarray}
 \frac{1}{D_{c}} &=& \frac{E_{c}}{L_{c}}=\sqrt{\frac{r_{c}-2M}{r_{c}^2(r_{c}-b)}}
\end{eqnarray}
Solving equation(\ref{phr1}), one could obtain the radius of the CPO:
\begin{eqnarray}
 (r_{c})_{\pm}&=& \frac{1}{4}\left(b+6M\pm \sqrt{b^2-20bM+36M^2} \right)
\end{eqnarray}
Here we can easily see that two situation arises first one is that $(r_{c})_{\pm}\geq 2M$ and the second one is $r_{c}<2M$.
Also $(r_{c})_{\pm}$ are real when $b\leq 2M$ or $b\geq 18M$. Since we are interested in this work the case for $b\leq 2M$,
therefore the radius of the CPO occurs at
\begin{eqnarray}
 r_{ph}&=&\frac{1}{4}\left(b+6M+\sqrt{b^2-20bM+36M^2} \right)
\end{eqnarray}

Again the angular frequency $\Omega_{c}$ measured by an asymptotic observers is given by
\begin{equation}
\Omega_{c}=\frac{u^\phi}{u^t}=\frac{1}{D_{c}}=\sqrt{\frac{r_{c}-2M}{r_{c}^2(r_{c}-b)}}
\end{equation}

\section{Extremal Case:}
Now we turn to the extremal cases to see what is happening there.

\subsection{Particle Orbits}

\begin{figure}[t]
\begin{center}
{\includegraphics[width=0.45\textwidth]{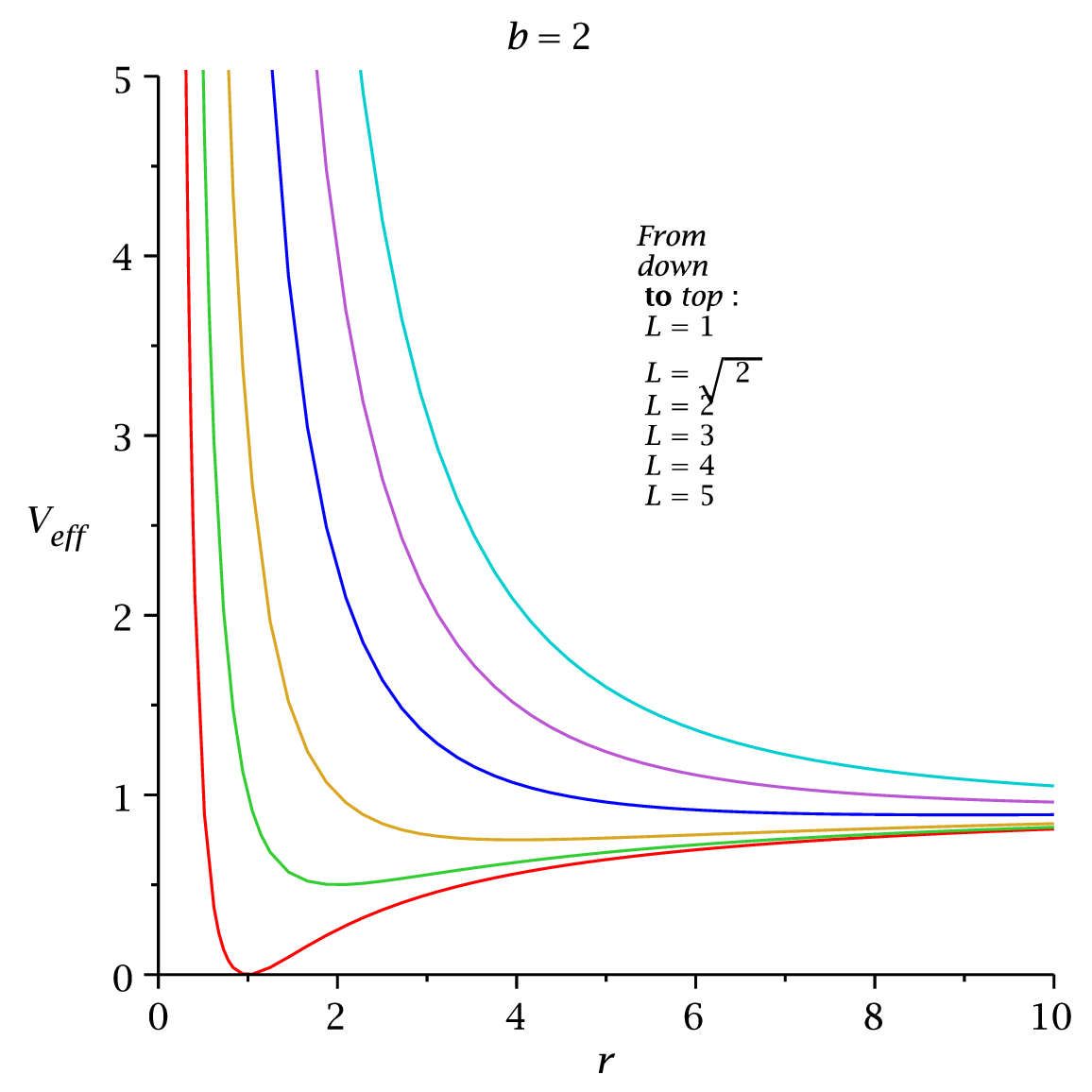}}
\end{center}
\caption{The picture shows the variation  of ${\cal V}_{eff}$  with $r$.  Here, $M =1$.
 \label{gmxcase}}
\end{figure}

Proceeding analogously, the corresponding effective potential  for
extremal GMGHS BH is found to be
\begin{eqnarray}
{\cal V}_{eff}=\left(1+\frac{L^{2}}{r(r-2M)} \right)\left(1-\frac{2M}{r}\right) ~.\label{vrn1}
\end{eqnarray}
It could be seen in the Fig. \ref{gmxcase}

\begin{figure}[t]
\begin{center}
{\includegraphics[width=0.45\textwidth]{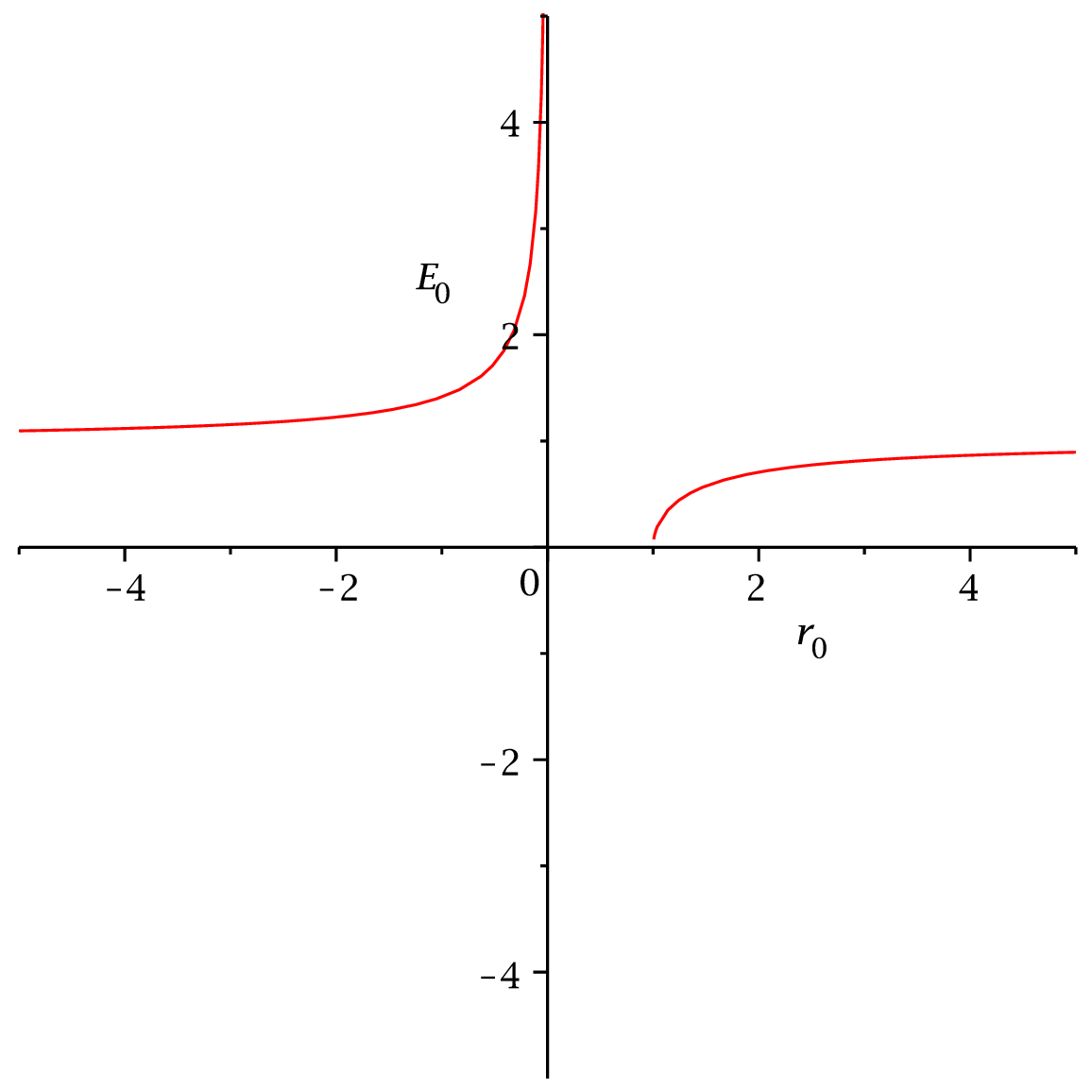}}
{\includegraphics[width=0.45\textwidth]{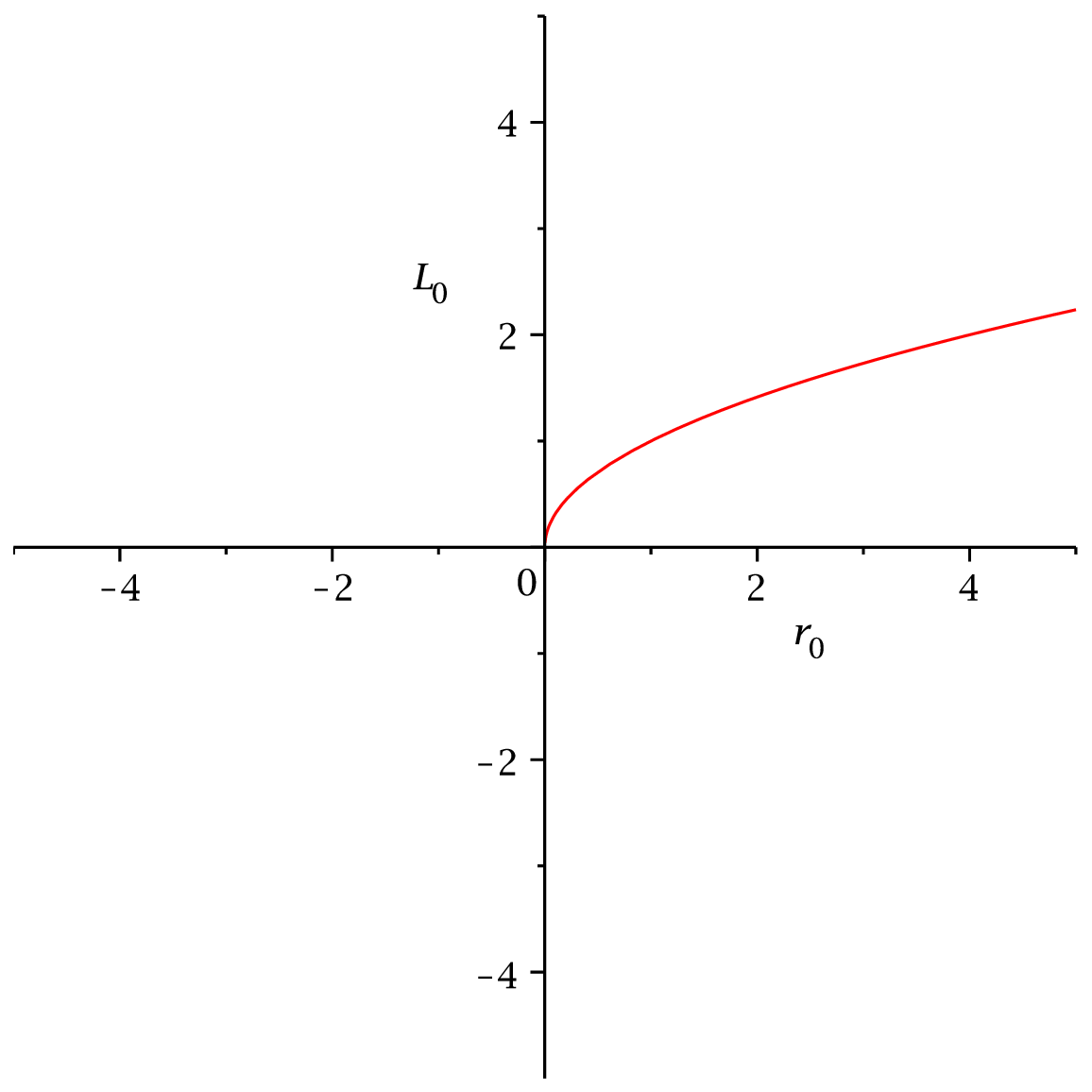}}
\end{center}
\caption{The figures display the variation of energy and  angular momentum with $r$
 \label{enagpcase}}
\end{figure}

Using the condition for circular geodesics of constant $r=r_{0}$,  we obtain the energy and angular
momentum per unit mass for the test particle as
\begin{eqnarray}
E_{0} &=& \sqrt{1-\frac{M}{r_{0}}} ~.\label{eng1}
\end{eqnarray}
and
\begin{eqnarray}
L_{0} &=&\sqrt{Mr_{0}}~ .\label{ang1}
\end{eqnarray}
In the Fig. \ref{enagpcase}, we have drawn the graph for energy and angular momentum.

Once again, for Circular geodesic motions of the test particle, both energy and angular momentum are real
and finite, therefore we must have $r_{0}>M$.
Now the most interesting class of circular orbit is the innermost stable circular orbit which occurs 
at the point of inflection as given by the equation (\ref{pi}). Hence the ISCO
equation for the test particle of the extremal GMGHS space-times can be written as

\begin{eqnarray}
 (r_{0}-2M)^3 &=& 0
\end{eqnarray}
Hence the ISCO occurs at the radius $r_{0}=r_{ISCO} =2M$ for extremal GMGHS BH.
At the ISCO the values of energy and angular momentum becomes $E_{ISCO} =\frac{1}{\sqrt{2}}$ and
$L_{ISCO}=\sqrt{2}M$ respectively.

\subsection{Photon Orbits:}

In this case,  the effective potential becomes
\begin{eqnarray}
 {\cal U}_{eff} &=& \frac{L^2}{r(r-2M)}\left(1-\frac{2M}{r}\right)
\end{eqnarray}

For circular geodesics of constant $r=r_{c}$, the  ratio of energy and angular momentum
\begin{eqnarray}
 \frac{E_{c}}{L_{c}} &=& \pm \frac{1}{r_{c}}
\end{eqnarray}
and
\begin{eqnarray}
 (r_{c}-2M)^2 &=& 0.~\label{phr}
\end{eqnarray}
After introducing the impact parameter $D_{c}=\frac{L_{c}}{E_{c}}$, the above equation reduces to
\begin{eqnarray}
 \frac{1}{D_{c}} &=& \frac{E_{c}}{L_{c}}=\frac{1}{r_{c}}
\end{eqnarray}
Solving equations(\ref{phr}) one could obtain  simply the radius of the circular photon orbit is
\begin{eqnarray}
 r_{c} &=& 2M
\end{eqnarray}
Therefore for extremal GMGHS BH, the photon orbit occurs at $r_{ph}=2M$. 

Therefore from the above analysis, we get for extremal GMGHS BH or extremal string BH
\begin{eqnarray}
r_{ISCO}=r_{ph}=r_{mb}=2M.~\label{rel1}
\end{eqnarray}
These three important geodesic orbits coincide with the event horizon. This is the key point of 
our investigation.

\section{Proper Radial Distance:}

Here we shall show that the proper spatial
distance on a spatial (constant time) slice, from an exterior point to
the horizon gives the zero value  for the extremal GMGHS BH.
The proper spatial distance \cite{sch} on a constant time slice from any point to the event
horizon is given by
\begin{eqnarray}
r_{p}&=&\int_{r_{hor}}^{r_{ISCO}}{\sqrt{g_{rr}}}\,\, dr\nonumber\\
     &=& \left(\sqrt{r(r-2M)}+2M\ln \mid {\sqrt{r}+\sqrt{r-2M}}\mid \right) \mid_{r_{hor}}^{r_{ISCO}}     \label{11}
\end{eqnarray}
where $g_{rr}=\frac{r}{r-2M}$.
In the near extremal limit, $Q=\sqrt{2}M(1-\chi)$. The event horizon is
located at $r_h = 2M$, the CPO is at $r_{ph}=M(2+2\sqrt{\chi}-\chi)$,
a MBCO is at $r_{mb}=2M(1+\sqrt{2\chi})$ and
the ISCO is at $r_{ISCO}=2M(1+\chi^{1/3}+2\chi^{2/3})$  for the equatorial plane. The
proper radial distance from photon orbit to event horizon at the extremal
limit $\chi \rightarrow 0$ is ${r_{p}}\mid
_{r_{hor}}^{r_{ph}}=0$. Again the
limiting distance from ISCO to the  horizon is ${r_{p}}\mid
_{r_{hor}}^{r_{ISCO}}=0$. The distance
for marginally bound orbit $(r_{mb})$ to the event horizon is ${r_{p}}\mid _{r_{hor}}^{r_{mb}}=0$,
which  is also vanishing in the extremal limit. Since all the proper radial distances
are vanishing at the extremal limit, therefore they must coincides with the
\emph{null generators of the horizon}. It may be noted that the metric components $g_{rr}$ is independent
of $Q$. Due to this fact, all the proper radial distances from horizon to any exterior point becomes zero.

\section{GMGHS space-time in Painlev\'{e}-Gullstrand Coordinates}

The standard discussion of ISCOs in previous section has given in terms of
Schwarzschild coordinates(SC)  which are known to be ill-behaved on the event
horizon. So in this section, we shall introduce a number of well behaved coordinate systems like ingoing
Eddington-Finklestein (EF), Painlev\'{e}-Gullstrand (PG)  coordinates which are regular on the event horizon
and  has some interesting properties. We shall show the effective potential derived in these coordinates are
similar to those obtained in SC.
Beginning with the  EF coordinates, we have the metric as
\begin{eqnarray}
ds^2=-\left(1-\frac{2M}{r}\right)dv^2+2dvdr+r(r-b)(d\theta^2+\sin^2\theta d\phi^2)\label{ef}
\end{eqnarray}
which is obtained by the following coordinate transformation for Schwarzschild BH
\begin{eqnarray}
v=t+r+2M \ln\mid \frac{r}{2M}-1\mid \label{vr1}
\end{eqnarray}
The above metric is the same as time independent, spherically symmetric geometry with different coordinates
and is not singular at $r=2M$. This type of coordinates are very useful to \emph{study the ongoing gravitational collapse}.
Now if we have given the following  transformation
\begin{eqnarray}
dv=dt+\frac{dr}{1+\sqrt{\frac{2M}{r}}}\label{dvr}
\end{eqnarray}
then we have found the well known metric of GMGHS BH
space-time in  Painlev\'{e} coordinates:
\begin{eqnarray}
ds^2=-\left(1-\frac{2M}{r}\right)dt^2+2\sqrt{\frac{2M}{r}}dtdr+dr^2+
r(r-b)(d\theta^2+\sin^2\theta d\phi^2) \label{3c}
\end{eqnarray}
which is unlike Schwarzschild coordinates, are not singular at the horizon. It is manifested that the space-time
now well behaved on the horizon. This type of coordinates could be used to calculate the Hawking radiation.

Similar to the Schwarzschild space-time, the GMGHS space-times also have isometries , namely time-like isometries
and rotational isometries as we have defined in the section II. Analogously, the radial equation in this coordinate
chart is found to be:
\begin{equation}
\left(1-f\right)\dot{r}^2+\left(1+\frac{L^2}{r(r-b)}\right)f={\cal E}^2 \label{3g}
\end{equation}
where $f=1-\frac{2M}{r}$.

For circular orbit $\dot{r}=\frac{dr}{d\tau}=0$ and by substituting the value of $f$ one obtains
\begin{eqnarray}
{\cal E}^2={\cal V}_{eff}=\left(1+\frac{L^2}{r(r-b)}\right)\left(1-\frac{2M}{r}\right) ~. \label{3h}
\end{eqnarray}
which is exactly similar to the effective potential as we have found in Schwarzschild coordinates.

Let us now study the properties of the
radial geodesics with zero angular momentum $L=0$ and radial free fall of a particle from infinity i.e. $E=1$ then
the Eq.(\ref{3g}) can be written as

\begin{eqnarray}
\frac{dr}{d\tau}&=& \pm 1 ~. \label{ff}
\end{eqnarray}
$+$ sign for outgoing geodesics and $-$ sign for ingoing geodesics. Now one can
compute the proper time interval which is found to be
\begin{eqnarray}
\tau &=& \int^{r_{ISCO}}_{r_{hor}}dr =r_{ISCO}-r_{hor} ~. \label{tau}
\end{eqnarray}
This implies that the proper time interval from event horizon to ISCO or event
horizon to MBCO or event horizon to CPO in the extremal limit gives
zero value. This is why, these three orbits coincide with the principal null generators 
of the horizon. In schwarzschild coordinates, the above calculation gives 
identical result also.

Now if we compare the Schwarzschild  time slice and Painlev\'{e}  time slice
for the GMGHS BH, we have
\begin{eqnarray}
ds^{2}_{Sch} &=& \frac{r}{r-2M}\, dr^{2}+r(r-b)d\phi^{2} \nonumber \\
ds^{2}_{Pain} &=&  dr^{2}+r(r-b)d\phi^{2}  ~. \label{3hh}
\end{eqnarray}
This implies  that in the extremal limit $b=2M$, on an equatorial constant time slice the proper radial
distance from horizon to any exterior point gives zero value, this means that in the extremal limit these
three orbits namely  ISCO, CPO and MBCO are precisely located on the
event horizon, which coincides with the null generators of the horizon.

\section{Discussion}
To summarize, we have demonstrated that the geodesic motion of a neutral test particle for time-like and null
circular geodesics in the equatorial plane of both extreme and non-extreme cases for GMGHS BH. We
have compared the ISCO for non-extremal BH as well as for extremal BH. The interesting findings for extremal 
BH is that, ISCO, CPO and MBCO coincides with the event horizon i.e. $r_{ISCO}=r_{ph}=r_{mb}=r_{hor}=2M$, which 
is quite different from extremal RN space-time.  Where the ISCO occurs at $r_{ISCO}=4M$ \cite{ms}, 
photon sphere \cite{cve} is located at $r_{ph} = 2M$, MBCO is situated at $r_{mb}=\frac{3+\sqrt{5}}{2} M$ and 
the horizon is at $r_{hor}=M$. Thus for this space-time, the inequality becomes 
$r_{ISCO}\neq r_{ph} \neq r_{mb} \neq r_{hor}=M$.

Since the proper radial distance on an equatorial constant time slice,
from  the ISCO to event horizon or photon orbit to event horizon or MBCO to event
horizon is exactly zero both in Schwarzschild and Painlev\'{e} coordinates, so they are in fact coalesce with
the principal null generators of the horizon. What is interesting in this space-time is that ISCO lies on the
unstable photon sphere. An another interesting feature of this space-time is that the area $8 \pi M(2M-b)$
goes to zero at the extremal limit, which is also quite different  from RN BH.

\section*{Acknowledgements}

I would like to thank Prof. P. Majumdar of R. M. V. U for helpful discussions. I also would like to thank the Editor 
Prof. J. P. S. Lemos and anonyomous referee for their helpful suggestions.

%\section*{References}

\end{document}